# Personal Moderation Configurations on Facebook: Exploring the Role of FoMO, Social Media Addiction, Norms, and Platform Trust

Shagun Jhaver

Personal moderation tools on social media platforms let users control their news feeds by configuring acceptable toxicity thresholds for their feed content or muting inappropriate accounts. This research examines how four critical psychosocial factors – fear of missing out (FoMO), social media addiction, subjective norms, and trust in moderation systems – shape Facebook users' configuration of these tools. Findings from a nationally representative sample of 1,061 participants show that FoMO and social media addiction make Facebook users more vulnerable to content-based harms by reducing their likelihood of adopting personal moderation tools to hide inappropriate posts. In contrast, descriptive and injunctive norms positively influence the use of these tools. Further, trust in Facebook's moderation systems also significantly affects users' engagement with personal moderation. This analysis highlights qualitatively different pathways through which FoMO and social media addiction make affected users disproportionately unsafe and offers design and policy solutions to address this challenge.

## Introduction

Social media sites empower users worldwide by letting them create and share the content of their choice. The rules determining acceptable content on these platforms often reflect the cultural norms predominant in Silicon Valley, where most major digital platforms are located (Gillespie, 2018). However, norms of appropriate conduct vary widely across different cultures and communities, so relying on a universal platform-driven approach to regulate online content overlooks the diverse requirements of millions of users (Chandrasekharan et al., 2018; Gorwa et al., 2020). Recognizing this, some scholars have called for an alternative approach, *personal moderation*, defined as a "form of moderation in which users can configure or customize some aspects of their moderation preferences on social media" (Jhaver et al., 2023).

Today, personal moderation tools are critical tools that social media platforms offer to let users choose their moderation preferences (Feng et al., 2024; Jhaver et al., 2023). Examples of these tools include *mute functionality*, which lets users stop an account from appearing in their news feed, and *word filters*, which allow users to configure a set of keywords they do not want to see on their feed. These tools empower users to align the content they view with their tolerance for sensitive spectatorship (Tait, 2008). Further, they help users avoid content-based harms, the ill effects of which have been extensively documented in prior literature (Jhaver et al., 2023; Roberts, 2019; Scheuerman et al., 2021).

Recognizing the utility of these tools, previous research has thus far analyzed how users configure them (Alqabandi et al., 2024; Feng et al., 2024; Jhaver and Zhang, 2023); designed and built novel personal moderation tools (Jhaver et al., 2022); and examined how changes in their design can produce more valuable outcomes for users, such as increased transparency about their decisions (Jhaver et al., 2023). However, we do not yet clearly understand what psychosocial factors regarding media use contribute to users' adoption of such tools in the first place. Given that these tools substantially shape user experience and safety on social media sites, it is vital that we understand the factors that affect their use.

I fill this critical gap by exploring how some key psychosocial factors regarding media use contribute to Facebook users' configuration of personal moderation tools. Building upon my findings from a nationally representative survey in the US, I discuss how *Fear of Missing Out (FoMO)* and *social media addiction* add new dimensions of vulnerability to online harms for the affected users. I also reflect on how platforms and communities can address this challenge by reinforcing *subjective norms* regarding the use of these tools and fostering *trust in content moderation systems*.

I chose these four factors, which I define and explain below, because they have been widely examined in prior literature regarding their relation to individuals' social media attitudes and behaviors, and they correlate with other factors of interest, e.g., prior research has linked FoMO to stress, depression, and anxiety (Elhai et al., 2016; Tugtekin et al., 2020). However, this is far from an exhaustive list, and many other factors (e.g., reciprocity, ease of use) may influence users' interactions with personal moderation tools. Therefore, I present these findings as a preliminary exploration of this significant issue.

## Background and Literature Review

### Content-based Harm, Online Safety, and Personal Moderation Tools

This research focuses on *content-based harm*, which refers to harm caused by viewing undesirable online content (Jhaver et al., 2022). Content-based harm has been documented across social media platforms like Reddit and Twitter (Sowles et al., 2018), video-sharing platforms like YouTube and TikTok (Lewis et al., 2012), gaming platforms like Twitch and Xbox, and elsewhere (Jhaver et al., 2018; Jhaver et al., 2022; Meisner, 2023). Prior research has shown that continuous exposure to violent, hateful, or otherwise troubling posts can negatively affect mental health, including inducing panic attacks and secondary trauma (Roberts, 2019).

Platforms usually address content-based harm with content moderation (Gillespie, 2010; Schoenebeck et al., 2021; Seering et al., 2020; West, 2018). When users post content that violates platform rules, platforms impose sanctions, such as removing that content or banning that user's account (Schoenebeck et al., 2021). Traditionally, those issuing sanctions may be commercial content moderators employed by the platform or community content moderators, who are volunteer end-users invested in their community's success (Seering et al., 2019). Previous studies have explored the limitations and challenges introduced by these existing strategies to mitigate content-based harms and promote online safety. For example, Pater et al. (2016) showed that online platforms often have inconsistent and non-exhaustive standards for addressing such harms. This can lead to frustration among targets who report content they believe to be in violation, only to discover later that the attack falls outside the scope of the remediation (Blackwell et al., 2017).

Some scholars have taken a victim-centric approach to addressing content-based harm, studying victims' experiences and perspectives on various aspects, including the classification of harm (Blackwell et al., 2017), the impact of harm (Scheuerman et al., 2021), and effective ways to address it (Jhaver et al., 2018). Xiao et al. (2023) found that current approaches fail to sufficiently remove disturbing materials afflicting victims, like fake news, alt-right trolls, and revenge porn, and further perpetuate harm by directing offenders' attention to the punishment they receive instead of the damage they cause. By examining the diverse needs of victims, researchers have explored interventions beyond platform-enacted content moderation. Examples include outsourcing the filtering of problematic content to friends (Mahar et al., 2018), providing tools to help victims gather authentic evidence of harm to share publicly (Sultana et al., 2021), or encouraging offenders to apologize to their victims (Xiao et al., 2023).

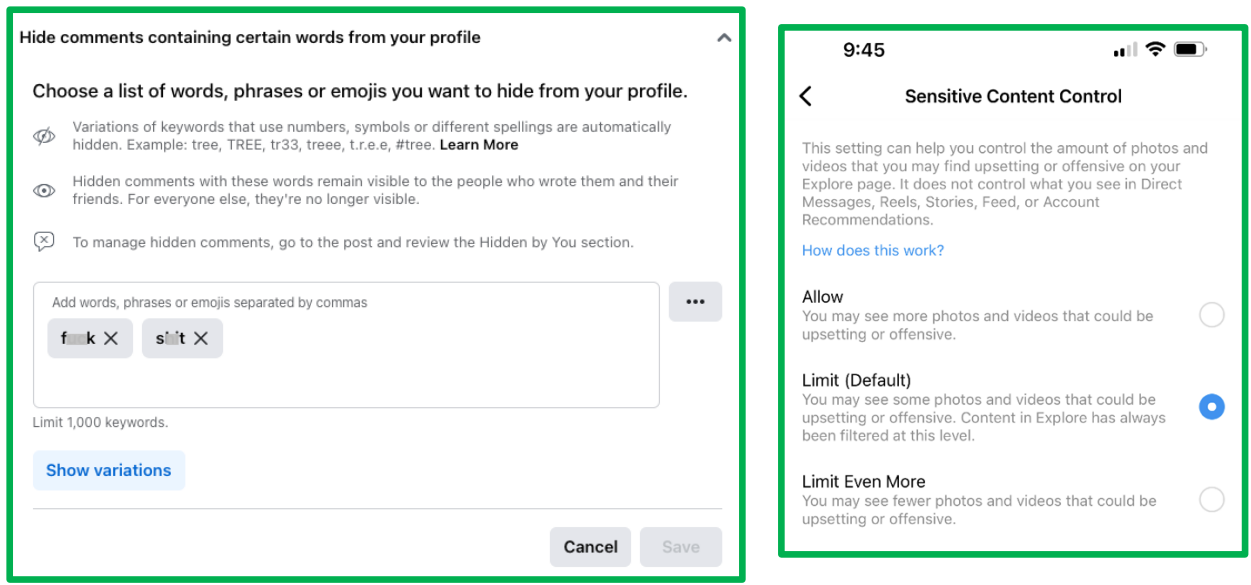

*Figure 1: Examples of personal content-based moderation tools on Facebook (left) and Instagram (right).*

I add to these prior efforts by focusing on the use of personal moderation tools that let users *proactively prevent* content-based harms (Jhaver et al., 2023). Prior inquiries on designing and deploying these tools have demonstrated that they can increase users' safety perceptions and enhance their participation online (Jhaver et al., 2018; Jhaver et al., 2023). A recent nationally representative survey of 984 US adults showed that end-users prefer personal moderation tools over platform-enacted moderation to prevent content-based harms resulting from exposure to hate speech, violent content, and sexually explicit content (Jhaver and Zhang, 2023). The current article advances this line of research by examining the psychosocial factors impacting users' propensity to configure these tools to reduce content-based harms.

Jhaver et al. (2023) classified personal moderation tools into two types: content-based and account-based. *Content-based tools* let users configure moderation preferences based on the content of each post (see Figure 1). For example, *word filter tools* allow any user to configure a set of undesired keywords (Figure 1, left); once set up, posts containing any of these keywords are automatically hidden from the user's news feed (Jhaver et al., 2022). Another category of content-based tools is *sensitivity controls* (Figure 1, right), which let users configure their moderation preferences on a Likert-type scale over factors like content sensitivity or toxicity (Jhaver et al., 2023). This study examines the use and configuration of *toxicity* sliders.

Unlike content-based tools, *account-based tools* (Figure 2) let users restrict their interaction with an individual account or a set of accounts (Geiger, 2016; Jhaver et al., 2018). For example, muting an account hides all subsequent posts from it to a user's news feed. This work examines users' preferences for muting inappropriate accounts.

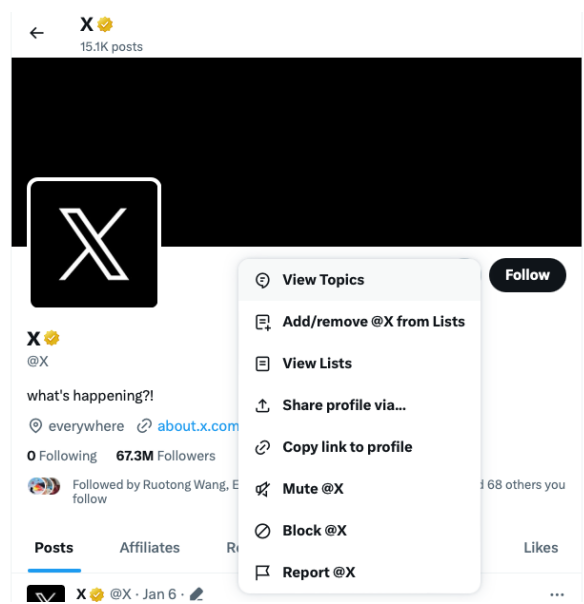

*Figure 2: Examples of account-based moderation actions available on Twitter. Users can choose to 'mute' or 'block' any account.*

## Fear of Missing Out (FoMO)

The Fear of Missing Out (FoMO) is defined as "a pervasive apprehension that others might be having rewarding experiences from which one is absent" (Przybylski et al., 2013). Błachnio and Przepiórka (2018) characterize FoMO as "a fundamental human motivation that consists in [of] craving interpersonal attachments." Such attachments can be impeded by social exclusion, which is frequently associated with the experience of social pain (Lai et al., 2016). FoMO arises due to temporary or persistent deficits in satisfying three fundamental psychological needs: competence, autonomy, and relatedness (Przybylski et al., 2013). Studies show that it is a widespread phenomenon, with 56% - 70% of adults suffering from FoMO (Murphy, 2013; Westin and Chiasson, 2021). Prior research has identified FoMO as a public health concern, linking it to stress, depression, anxiety (Elhai et al., 2016; Tugtekin et al., 2020), headaches (Baker et al., 2016), and decreased sleep (Milyavskaya et al., 2018).

While FoMO was initially conceptualized in the offline context (Przybylski et al., 2013), it has found widespread applicability regarding social media use (Bloemen and De Coninck, 2020; Reer et al., 2019). Over the past decade, researchers have established its evidentiary relationship with online vulnerability (Thompson et al., 2021), fake news and misinformation sharing (Talwar et al., 2019), and social media fatigue (Malik et al., 2020). Some researchers have also examined the connection between platform users' FoMO and their interactions with system design. For example, Westin and Chiasson (2021) showed that social media users are systematically pressured into privacy-compromising behaviors, such as posting more information more often, due to FoMO. Popovac and Hadlington (2020) demonstrated that FoMO is a significant predictor of online risk-taking behaviors, such as sexting and sharing passwords with friends, among adolescents. I build upon this research to examine how FoMO influences users' online safety practices. Specifically, I explore how FoMO affects users' configuration of personal moderation tools that proactively reduce online harm.

## Social Media Addiction

Social media addiction refers to "the irrational and excessive use of social media to the extent that it interferes with other aspects of daily life" (Hou et al., 2019). Prior literature has shown that users suffering from social media addiction exhibit most behavioral addiction symptoms, including tolerance, withdrawal, conflict, salience, relapse, and mood modification (Andreassen, 2015; Griffiths et al., 2014; Ryan et al., 2014). Social media addiction is associated with greater loneliness, anxiety, and suicidality and with declines in academic performance, self-esteem, and life satisfaction (Hawi and Samaha, 2016; Latikka et al., 2022). Increased social media use also raises individuals' potential exposure to various online vulnerabilities, including online harassment, incidents of data misuse, interactions with strangers with harmful intentions, and exposure to inappropriate content due to spending more time on these sites (Brandtzæg et al., 2010; Staksrud et al., 2013; Sasson and Mesch, 2014).

Contemporary social theories offer frameworks to explain the links between social media addiction and safety practices (Craig et al., 2020). According to Problem Behavior Theory (Jessor and Jessor, 1977), risky behaviors tend to co-occur, and certain individuals have distinct characteristics that heighten their susceptibility to engaging in such behaviors (Craig et al., 2020). For example, Huang et al. (2023) demonstrated a pathway between social media addiction and food addiction with the involvement of psychological distress. I examine whether social media addiction relates to another risky behavior: configuring inadequate moderation controls. Further, drawing from social learning theory, individuals who spend more time on social media may observe more offensive behaviors and, through role modeling and reinforcement, may come to see them as more acceptable (Lee et al., 2023). Repeated exposure to offensive behaviors may also result in the "disinhibition effect" (Barlett and Gentile, 2012), i.e., inappropriate behaviors may become more normalized over time. This suggests that users with social media addiction may see offensive content as acceptable and hesitate to proactively remove it from their feeds. To examine this, I study how social media addiction influences users' choices regarding personal moderation tools.

## Subjective Norms

According to Ajzen (1991), subjective norms refer to the social pressure associated with performing a given behavior. This pressure may stem from two categories of subjective norms: descriptive and injunctive norms (Ajzen, 2020). Descriptive norms are beliefs about whether important others themselves perform the behavior under consideration, whereas injunctive norms are expectations about whether others approve or disapprove of that behavior (Ajzen, 2020; Kiesler et al., 2012). The underlying assumption is that people generally engage in behaviors that are encouraged and embraced within their social sphere.

Prior research has reported that subjective norms significantly affect individuals' behaviors in a variety of social media contexts, such as posting behaviors (Arpaci, 2020), privacy regulation (Neubaum et al., 2023), and preventing unruly conversations (Matias, 2019). Subjective norms also contribute to engagement in negative behaviors, such as taking risky selfies (Chen et al., 2019), excusing the use of aggressive language (Allison et al., 2019), and conducting cyberbullying (Heirman and Walrave, 2012). Recent literature has specifically acknowledged the role of subjective norms in people's responses to fake news, health misinformation, and conspiratorial content (Colliander, 2019; Koo et al., 2021; Bautista et al., 2022). Building upon this literature, I examine how subjective norms shape users' preferences for addressing content-based harms through configuring personal moderation tools.

## Trust in the Moderation System

Personal moderation tools are automated tools whose operations rely on the efficacy of algorithmic mechanisms that drive them. Users of any AI application must have the confidence that they can depend on the AI agent to accomplish their objectives in situations of uncertainty (Okamura and Yamada, 2020). The "Computers are Social Actors" paradigm suggests that viewing AI applications as human-like collaborators rather than tools can clarify our understanding of human trust in AI (Seeber et al., 2020). Prior literature on user acceptance of AI systems emphasizes interpersonal trust as an essential component (Gillath et al., 2021); such trust encompasses the willingness of one party to accept vulnerability based upon positive expectations of the behavior of another party, irrespective of the ability to monitor that party (Lewicki et al., 2006).

Accordingly, I conceptualize trust in personal moderation tools as the extent to which a user is confident and comfortable in the actions or decisions made by these tools and is therefore willing to rely on them. This includes faith in the judgment and fairness of the platform's moderation system to determine the appropriateness of submitted posts and confidence that these tools' decisions would align with users' own determinations.

Recent controversies over Facebook's privacy invasions have affected users' trust in its moderation system (Brown, 2020). This could, in turn, influence the use of the platform's safety tools. Prior research has examined how users' trust in algorithmic moderation tools shapes their perspectives about content moderation decisions (Jhaver et al., 2019; Schulenberg et al., 2023; Molina and Sundar). I build upon this research to examine the extent to which users' trust in the moderation apparatus affects their configuration of personal moderation tools.

## Methods

For this study, which was considered exempt from review by the Institute's[i] IRB, I recruited participants via Lucid,[ii] a survey company that provides researchers access to demographically representative national samples. This survey's inclusion criteria encompassed all adult internet users within the United States. Compensation for participants was facilitated through the Lucid system.

I framed this survey's questionnaire around Facebook because its widespread popularity made it more likely that many users would be aware of terms relevant to the concepts this study explores, such as 'Facebook friends' and 'news feeds.' By focusing on Facebook, this paper also adds to vital research (Paradise and Sullivan, 2012; Mena, 2020; Théro and Vincent, 2022; Alhabash and Ma, 2017) into this key social network site's (actual and perceived) social implications and use.

I designed survey questions to examine how four key psychosocial factors related to media use – FoMO, social media addiction, subjective norms, and trust in moderation systems – shape Facebook users' adoption or configuration of two personal moderation tools: (1) sensitivity controls and (2) the muting function. In doing so, I adapted survey instruments from pertinent prior research to evaluate specific measures, which I explain below. To enhance the survey's validity, I solicited input on early versions of the questionnaire from peers and students within my institution. Eight individuals external to the project, trained in diverse fields like Computer Science, Psychology, and Media Studies, offered responses and suggestions regarding question phrasing and survey structure. I integrated these insights into the survey design. Next, I conducted a trial run of the survey with 30 participants from Lucid, which prompted further refinement of the questionnaire.

I conducted the survey through the online survey platform Qualtrics, with the survey going live on Jan 4, 2024. Since this survey focuses on Facebook use, I screened for participants who had used Facebook over the past year. I also implemented data-cleaning steps to improve the quality of analyzed survey responses. For example, I removed responses where participants engaged in straightlining (Kim et al., 2018), i.e., selecting identical answers (e.g., 'Strongly agree') to items in every question that uses the same response scale. Table 1 shows the demographic details of my final sample, which comprised 1,061 participants following data refinement. This table also compares the demographics of my sample to those of adult internet users in the United States.

*Table 1: Demographic details of survey participants*

|  | **This study, US Survey** Jan 2024 (%) | **American Community Survey**, US sample 2021 (%) |
|---|---|---|
| **Age** |  |  |
| 18-29 | 13.3 | 17.4 |
| 30-49 | 39.6 | 29.5 |
| 50-64 | 24.9 | 25.6 |
| 65+ | 22.2 | 27.3 |
| **Gender** |  |  |
| Male | 48.3 | 48.6 |
| Female | 51.7 | 51.4 |
| **Race/Ethnicity** |  |  |
| White | 74.8 | 68.3 |
| Black | 12.6 | 9.3 |
| Other | 12.6 | 22.4 |
| **Hispanic** |  |  |
| Yes | 11.3 | 13.7 |
| **Education** |  |  |
| High school or less | 27.5 | 33.5 |
| Some college | 34.9 | 33.3 |
| College+ | 37.6 | 33.1 |

# Measures

I built hierarchical linear regression models to examine the relationships between this study's independent variables (FoMO, Facebook addiction, subjective norms (injunctive and descriptive), and trust in moderation) and dependent variables (sensitivity control setting and likelihood of muting). I describe below how I used survey items to measure each variable along with the variable's mean (M), standard deviation (SD) and Cronbach's alpha ($\alpha$) values from the processed survey data. I also note the socio-demographic variables I controlled for in these models.

## Fear of Missing Out (FoMO)

Participants responded to the ten-item Fear of Missing Out scale (Przybylski et al., 2013) with answer choices ranging from 1 = "Not at all true of me" to 5 = "Extremely true of me." This scale measures the extent of apprehension of missing out on the rewarding experiences of others. Example items include, "When I miss out on a planned get-together, it bothers me." and "I get worried when I find out my friends are having fun without me." The scale produced an average score ranging from 1 to 5, with higher scores indicating increased levels of FoMO. Consistent with past research (Przybylski et al., 2013), this FoMO scale showed good internal consistency (M = 2.01, SD = .73, $\alpha$ = .85).

## Facebook Addiction

Facebook addiction was measured using the 6-item Facebook Bergen Addiction Scale (FBAS) (Andreassen et al., 2012). Participants were prompted to "Please answer the following questions with regard to your Facebook use over the past year." I retained the use of 'Facebook' in these items because the rest of the survey also referred to Facebook as the site of focus. Each item in the FBAS corresponds to one of the six central components of addiction according to the Griffiths et al. (2014) model: salience, mood modification, tolerance, withdrawal symptoms, conflict, and relapse. For example, the item concerning withdrawal asked, "Have you become restless or troubled if you have been prohibited from using Facebook?" Participants responded on a 5-point Likert scale ranging from 1 = "Very rarely" to 5 = "Very often." These items were averaged to create a single measure and showed good internal consistency (M = 2.03, SD = .84, $\alpha$ = .84).

## Subjective Norms

Subjective norms regarding the configuration of personal moderation tools were assessed based on items adopted from prior research (Bautista et al., 2022; Pundir et al., 2021). Participants rated each of these items on a Likert-type scale ranging from 1 (Strongly disagree) to 7 (Strongly agree).

Injunctive norms regarding sensitivity controls and the muting function were assessed by the following items, respectively: (a) "Suppose Facebook offers a moderation setting that limits the number of upsetting or offensive posts appearing on my news feed. Most people who are important to me would expect me to turn on this setting," and (b) "Most people who are important to me would expect me to mute a Facebook friend's account if it frequently posts upsetting or offensive content." These two items were averaged to create an index for injunctive norms (M = 3.84, SD = 1.66, $\alpha$ = .74)

Similarly, descriptive norms regarding sensitivity controls and the muting function were assessed by the following items, respectively: (a) "Suppose Facebook offers a moderation setting that limits the number of upsetting or offensive posts appearing on the news feed. Most people who are important to me would turn on this setting in their Facebook profile," and (b) "Most people who are important to me would mute a Facebook friend's account if it frequently posted upsetting or offensive content." Averaging these items created an index for descriptive norms (M = 4.31, SD = 1.53, $\alpha$ = .78).

## Trust in the Moderation Process

This variable was operationalized using four items, each on a Likert-type scale ranging from 1 (Strongly disagree) to 7 (Strongly agree). First, faith in the judgment and fairness of the platform's moderation system was assessed by the following items: (a) "Facebook's content moderation process can be trusted to judge how upsetting or offensive each post is," and (b) "Facebook's content moderation process is fair and impartial in determining how upsetting or offensive each post is." Second, the expected alignment of users' content evaluations with Facebook's moderation was measured by asking: "If Facebook's content moderation process determines that a post is upsetting or offensive, I am likely to find it upsetting or offensive." Finally, users' confidence in Facebook's account-based moderation tools operating as expected was measured by asking: "I feel confident that if I mute an account on Facebook, its posts will no longer appear on my news feed." These four items were averaged to create a measure of general trust in Facebook moderation (M = 3.91, SD = 1.25, $\alpha$ = .77).

> Suppose that Facebook provides a setting that lets you control the volume of posts that you may find upsetting or offensive on your news feed. How would you configure this setting?
>
> ○ **Allow:** You may see more posts that could be upsetting or offensive.
> ○ **Limit:** You may see some posts that could be upsetting or offensive.
> ○ **Limit even more:** You may see fewer posts that could be upsetting or offensive.

*Figure 3: Survey question asking participants to configure their preference for a hypothetical content-based moderation tool on Facebook.*

## Dependent Variables Related to Personal Moderation

The adoption of personal moderation tools was assessed using two items. First, I asked for sensitivity controls: "Suppose that Facebook provides a setting that lets you control the volume of posts you may find upsetting or offensive on your news feed. How would you configure this setting?" Options included: (a) Allow, (b) Limit, and (c) Limit even more (Figure 3). This design was inspired by a similar setting offered by Instagram (Figure 1). Second, I asked for account-based tools: (a) "Suppose that you encounter a Facebook friend frequently posting upsetting or offensive content that appears on your news feed. How likely are you to mute this friend's account?" I added a note with this question that explained how "muting" works. This item was rated on a 7-point Likert-type scale ranging from 1 (Extremely unlikely) to 7 (Extremely likely).

## Control Variables

Prior literature demonstrated the relationship between socio-demographic variables and attitudes toward media regulation (Gunther, 2006; Lambe, 2002; Jhaver and Zhang, 2023). Therefore, I controlled for age, education, gender, race, and political affiliation (1 = "very liberal," 7 = "very conservative").

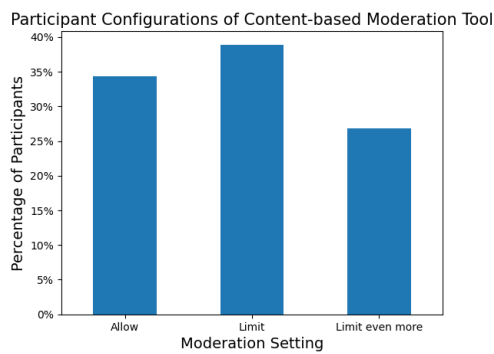

*Figure 4: Participants' configurations of sensitivity controls.*

# Results

Results show that 34.3%, 38.9%, and 26.8% of participants would prefer to configure their content-based moderation setting to 'Allow', 'Limit', and 'Limit even more' levels, respectively (Figure 4). Further, 29.5% of participants are at least slightly unlikely to mute an offensive Facebook friend, whereas 47.8% of participants are at least slightly likely to mute in such a case (Figure 5).

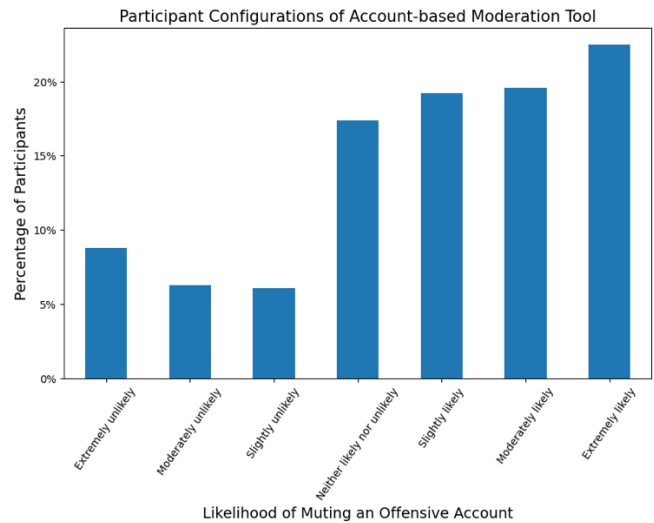

*Figure 5: Participants' configurations of the muting function.*

## Support for Stricter Content-based Personal Moderation

I computed hierarchical linear regression to test how different factors shape preferences for configuring Facebook's sensitivity controls. I created a model in which the dependent variable was the strictness of this tool setting that participants selected. In Step 1, I included the control variables age, gender, race, education, and political affiliation. In Step 2, I introduced five independent variables: (1) FoMO, (2) Facebook addiction, (3) Injunctive norms, (4) Descriptive norms, and (5) Trust in Facebook moderation (*Table 2*).

The regression model (Model 1) shows a significant negative influence of Facebook addiction on preference for setting up stricter sensitivity controls ($\beta$ = -.081, $p$ < .05). Further, injunctive norms ($\beta$ = .174, $p$ < .001), descriptive norms ($\beta$ = .207, $p$ < .001), and trust in Facebook moderation ($\beta$ = .145, $p$ < .001) all positively influence a preference for configuring a stricter setting. FoMO did not have a significant influence on participants' configurations of this content-based moderation tool.

*Table 2: Hierarchical multiple regression analyses predicting participants' preferences for setting up account- and content-based personal moderation tools (N = 1,061).*

| Independent Variable | Support for setting stricter sensitivity ($\beta$) | Likelihood of muting offensive account ($\beta$) |
|---|---|---|
| Model # | Model 1 | Model 2 |
| Step 1 | | |
|   Age | .122*** | .081** |
| Gender (Female) | .156*** | .104*** |
| Race (White) | -.097** | .000 |
| Education[a] | -.006 | .065* |
| Political affiliation[b] | .012 | -.038 |
| $R^2$ | .062*** | .044*** |
| Step 2 |  |  |
| Fear of missing out (FoMO) | -.053 | -.095** |
| Facebook addiction | -.081* | .063 |
| Injunctive norms | .174*** | .158*** |
| Descriptive norms | .207*** | .282*** |
| Trust in Facebook moderation | .145*** | .134*** |
| $R^2$ change | .160*** | .222*** |
| Total $R^2$ | .222*** | .266*** |

*p < .05, **$p$ < .01, ***$p$ < .001 ($t$ test for $\beta$, two-tailed; F test for $R^2$, two-tailed).
[a]0= Less than secondary education; 1= Secondary education or more.
[b]1= Strong Democrat, 7= Strong Republican.
[c]1= Less than 10 minutes per day, 6= More than 3 hours per day.

$\beta$ = Standardized beta from the full model (final beta controlling for all variables in the model).

## Likelihood of Engaging in Account-based Personal Moderation

Next, I computed hierarchical linear regression to test how different factors shape participants' likelihood of muting an account that frequently posts upsetting or offensive content. I created a model in which the dependent variable was the likelihood of muting such an account. As in Model 1, in Step 1, I included the control variables age, gender, race, education, and political affiliation. In Step 2, I introduced five independent variables: (1) FoMO, (2) Facebook addiction, (3) Injunctive norms, (4) Descriptive norms, and (5) Trust in Facebook moderation (*Table 2*).

This regression model (Model 2) shows a significant negative influence of fear of missing out (FoMO) on the likelihood of muting ($\beta$ = -.095, $p$ < .01). Additionally, injunctive norms ($\beta$ = .158, $p$ < .001), descriptive norms ($\beta$ = .282, $p$ < .001), and trust in Facebook moderation ($\beta$ = .134, $p$ < .001) all positively influence the muting action. Facebook addiction did not influence participants' likelihood of muting.

## Discussion

Personal moderation tools help users shape their social media feeds according to their preferences. These tools are one of the key ways in which users exert their voice in online governance infrastructures. Further, unlike reporting mechanisms (Crawford and Gillespie, 2016), these tools affect only the configuring account's feeds. Therefore, they present an opportunity for users to personalize their feeds without concern for how their actions affect other users. Prior research has also documented the utility of these tools in preventing exposure to content-based harms, such as hate speech and violence (Feng et al., 2024; Jhaver et al., 2023; Jhaver and Zhang, 2023). However, this literature does not relate how the use of these tools is influenced by various psychosocial factors. This paper begins to fill this vital gap.

My analysis shows that the fear of missing out (FoMO) reduces the likelihood of muting an account frequently posting offensive or upsetting content. When considering whether to mute an account, users make a tradeoff: should they avoid exposure to harmful content from that account or risk losing access to relevant content posted by that account? A significant relationship between FoMO and muting behavior identified here is a theoretically significant finding: it suggests that *FoMO makes users more vulnerable to content-based harms by deterring in-the-moment actions against violating accounts*. This adds another layer to previous findings on how FoMO contributes to online safety practices, such as privacy-compromising and risk-taking behaviors (Westin and Chiasson, 2021; Popovac and Hadlington, 2020).

I also found that *social media addiction significantly reduces the strictness levels users select for their sensitivity controls*. As prior research (Brandtzæg et al., 2010; Staksrud et al., 2013; Sasson and Mesch, 2014) points out, social media addiction raises exposure to inappropriate content just because the addicted users spend more time on these sites. My analysis shows that these users' vulnerability is further increased by their hesitance to configure stricter controls in moderation toggles. It is likely that, as Barlett and Gentile (2012) observe, repeated exposure to inappropriate content normalizes it for addicted users, and this disinhibition affects their moderation configurations. Thus, *social media addiction creates a vicious cycle whereby affected users are disproportionately exposed to online harms and become less likely to take actions that reduce this exposure.* Therefore, it is vital that design and policy efforts focus on reducing the prevalence of social media addiction. For instance, platforms could offer personalized psychological and mental health information about social media usage to heavy users, and prompt them to set and track their site usage goals (Cham et al., 2019).

How can platforms encourage users to engage in safety practices grounded in personal moderation choices? My analysis shows that both descriptive and injunctive norms significantly influence users to set up stricter settings in moderation toggles and mute inappropriate accounts. This finding opens *a design space for platforms to incorporate normative information in their sites as a key strategy to promote online safety*. For example, platforms may show users aggregated descriptive statistics of how their linked accounts (e.g., Facebook "friends" or Twitter "followees") set up their sensitivity controls. These information nudges may normalize and enhance the use of personal moderation tools and reduce users' exposure to content-based harms. This finding also suggests that users and communities can encourage the adoption of personal moderation tools among their social connections as an online safety mechanism by rendering the norms about their use more salient, e.g., by sharing how they help reduce content-based harm.

My findings also reveal that *trust in Facebook moderation significantly influences users' adoption of stricter content-based moderation toggles and their muting of inappropriate accounts.* Recent controversies about social media moderation decisions (Brown, 2020) and increased scrutiny of moderation infrastructures by news media, lawmakers, and scholars have diluted the general public's trust in social media platforms (Gillespie, 2018). The relationship between this trust and users' moderation practices documented here further emphasizes the importance of platform investment in gaining users' trust. It is vital that platforms take concrete and conspicuous actions to clarify their commitment to user safety and highlight the robustness of their content moderation mechanisms.

## Limitations and Future Work

The survey method deployed in this study was not intended to explore participants' specific motivations behind their different preferences for personal moderation. Further, in this cross-sectional study, definitive conclusions about causal relationships cannot be drawn. Instead, my investigation serves as an initial exploration of how some of the most relevant psychosocial factors about social media use identified in prior literature influence users' moderation actions. Future research can build upon this work by delving into the specific pathways through which FoMO and Facebook addiction affect participants' moderation configurations.

Finally, the survey questions were deliberately tied to a specific platform, Facebook, to ensure that participants could better comprehend and more concretely answer survey questions about their social media activities and preferences. Future studies should evaluate how these findings apply to other social media platforms.

## Conclusion

This study shows how FoMO and social media addiction decrease users' inclination to engage in proactive approaches to reducing content-based harm. Thus, it highlights a qualitatively different pathway, i.e., greater exposure to upsetting content, through which users experiencing FoMO and social media addiction could be disproportionately vulnerable to online harm. This paper also explores new directions for how platforms and community members can help such users. First, my findings suggest that platforms should offer these tools by default, and end-users and communities should normalize their use. Second, this study motivates design and policy efforts to reduce FoMO and social media addiction. Third, it demonstrates how user trust in content moderation systems plays a crucial role in their adoption of defensive tools and motivates further acceleration of efforts to foster this trust.

---

[i] I will specify the Institute name after the peer review process is completed.
[ii] https://lucidtheorem.com